\documentstyle[sprocl,epsf,cite]{article}

\def\Journal#1#2#3#4{{#1} {\bf #2}, #3 (#4)}

\def\NPB{{\em Nucl. Phys.} B}
\def\PLB{{\em Phys. Lett.}  B}
\def\PRL{\em Phys. Rev. Lett.}
\def\PRD{{\em Phys. Rev.} D}
\def\ZPC{{\em Z. Phys.} C}
\def\MPL{{\em Mod. Phys. Lett.} A}
\def\PR{\em Phys. Rep.}
\def\PTP{\em Prog. Theor. Phys.}

\def\be{\begin{equation}}
\def\ee{\end{equation}}
\def\bea{\begin{eqnarray}}
\def\eea{\end{eqnarray}}
\newcommand{\gev}{\,{\rm GeV}}
\newcommand{\tev}{\,{\rm TeV}}
\newcommand{\lsim}{
  \raisebox{0.2em}{$<$} \hspace{-0.75em} \raisebox{-0.2em}{$\sim$} }
\newcommand{\gsim}{
  \raisebox{0.2em}{$>$} \hspace{-0.75em} \raisebox{-0.2em}{$\sim$} }
\begin{document}

\title{New constraint on the minimal SUSY GUT model from proton 
decay\footnote{Talk given by T. Nihei in International Symposium on 
Supersymmetry, Supergravity, and Superstring (SSS99), Seoul, Korea, 
June 23-27, 1999, based on the published work \cite{GN}.}}

\author{Toru Goto and Takeshi Nihei}

\address{Theory Group, KEK, Tsukuba, Ibaraki 305-0801, Japan} 


\maketitle\abstracts{We present results of reanalysis
on proton decay in the minimal SU(5) SUSY GUT model. Unlike previous 
analyses, we take into account a Higgsino dressing diagram of 
dimension 5 operator with right-handed matter fields ($RRRR$ operator). 
It is shown that this diagram gives a significant contribution 
for $p \rightarrow K^{+}\overline{\nu}_{\tau}$. Constraints on the 
colored Higgs mass $M_C$ and the sfermion mass $m_{\tilde{f}}$ from 
Super-Kamiokande limit become considerably stronger than those in the 
previous analyses: 
$M_C$ $>$ $6.5 \times 10^{16} \gev$ for $m_{\tilde{f}}$ $<$ $1 \tev$.
The minimal model with $m_{\tilde{f}}$ $\lsim$ $2 \tev$ 
is excluded if we require the validity of this model up to the Planck scale. }
%
%%%%%%%%%%%%%%%%%%%%%%%%%%%%%%%%%%%%%%%%%%%%%%%%%%%%%%%%%%%%%%%%%%%%%%%%%%
%
\section{Introduction}
Supersymmetric grand unified theory (SUSY GUT) \cite{SUSY_GUT} 
is strongly suggested by gauge coupling unification 
around $M_X \sim 2 \times 10^{16} \gev$ \cite{Gauge_Coupling_Unification}. 
In this theory, the hierarchy between the weak scale and the GUT scale
$M_X$ is protected against radiative corrections by supersymmetry. 
Also, this theory makes successful prediction for the charge quantization. 
Proton decay is one of the direct consequences of grand unification. 
The main decay mode $ p \rightarrow K^{+}\overline{\nu} $ \cite{DRW} 
in the minimal SU(5) SUSY GUT model \cite{SUGRA} 
has been searched for with underground experiments \cite{Kam,IMB},
and the previous results have already imposed severe
constraints on this model. 
Recently new results of the proton decay search at 
Super-Kamiokande have been reported \cite{superK}. The bound on the
partial lifetime of the $K^+\overline{\nu}$ mode is
 $\tau(p\rightarrow K^+ \overline{\nu})$
$>$ $5.5 \times 10^{32}$ years (90\,\% C.L.)\footnote{After we finished
our analysis, the latest limit of the Super-Kamiokande 
$\tau(p\rightarrow K^+ \overline{\nu})$ $>$ $6.7 \times 10^{32}$ years 
(90\,\% C.L.) \cite{superK-new} was reported. 
An adaptation to the updated experimental limit is straightforward 
(See Eq.(\ref{eqn:constraint})).},
where three neutrinos are not distinguished.

There are many analyses on the nucleon decay in the minimal SU(5) 
SUSY GUT model \cite{dim5_op,DRW,NCA,MATS+HMY,HMTY,GNA}.
In the previous analyses, it was considered that the contribution 
from the dimension 5 operator with left-handed matter 
fields ($LLLL$ operator) was dominant 
for $ p \rightarrow K^{+} \overline{\nu} $ \cite{DRW}. 
In particular a Higgsino dressing diagram of the $RRRR$ operator 
has been estimated to be small or neglected in these analyses. 
It has been concluded that the main decay mode is 
$ p \rightarrow K^{+} \overline{\nu}_\mu $, 
and the decay rate of this mode can be suppressed sufficiently 
by adjusting relative phases between Yukawa couplings
at colored Higgs interactions \cite{NCA}.

In this talk, we present results of our analysis on the proton decay 
including the $RRRR$ operator in the minimal SU(5) SUSY GUT model 
\cite{GN}\footnote{See also Ref.\cite{Babu-Strassler}.}.
We calculate all the dressing diagrams \cite{NCA} (exchanging
the charginos, the neutralinos and the gluino) of 
the $LLLL$ and $RRRR$ operators, 
taking account of various mixing effects among the SUSY particles, 
such as flavor mixing of quarks and squarks, left-right mixing of 
squarks and sleptons, and gaugino-Higgsino mixing of charginos 
and neutralinos. 
For this purpose we diagonalize mass matrices numerically to
obtain the mixing factors at `ino' vertices 
and the dimension 5 couplings. 
Comparing our calculation with the Super-Kamiokande limit, 
we derive constraints on the colored Higgs mass
and the typical mass scale of squarks and sleptons. 
We find that these constraints are much stronger than 
those derived from the analysis neglecting the $RRRR$ effect.

%
%%%%%%%%%%%%%%%%%%%%%%%%%%%%%%%%%%%%%%%%%%%%%%%%%%%%%%%%%%%%%%%%%%%%%%%%%%
%
\section{Dimension 5 operators in the minimal SU(5) SUSY GUT model}
Nucleon decay in the minimal SU(5) SUSY GUT model is 
dominantly caused by dimension 5 operators \cite{dim5_op}, which 
are generated by the exchange of the colored Higgs multiplet. 
The dimension 5 operators relevant to the nucleon decay are described 
by the following superpotential: 
\begin{eqnarray}
W_5 & = & -\frac{1}{M_C} \left\{ \frac{1}{2}C_{5L}^{ijkl}
Q_k Q_l Q_i L_j + C_{5R}^{ijkl} U^c_i D^c_j U^c_k E^c_l \right\}.
\label{eqn:dim5_op}
\end{eqnarray}
Here $Q$, $U^c$ and $E^c$ are chiral superfields which contain 
a left-handed quark doublet, 
a charge conjugation of a right-handed up-type quark, 
and a charge conjugation of a right-handed charged lepton, 
respectively, and are embedded in the 10 representation of SU(5). 
The chiral superfields $L$ and $D^c$ contain a left-handed lepton 
doublet and a charge conjugation of a right-handed down-type quark,
respectively, and are embedded in the $\overline{5}$ representation. 
A mass of the colored Higgs superfields is denoted by $M_C$. 
The indices $ i,j,k,l = 1,2,3$ are generation labels. 
The first term in Eq.~(\ref{eqn:dim5_op}) represents the $LLLL$ 
operator \cite{DRW} which contains only left-handed SU(2) doublets. 
The second term in Eq.~(\ref{eqn:dim5_op}) represents the $RRRR$ 
operator which contains only right-handed SU(2) singlets.
%
%-------------
The coefficients $C_{5L}$ and $C_{5R}$ in Eq.~(\ref{eqn:dim5_op}) at the
GUT scale are determined by Yukawa coupling matrices as \cite{NCA} 
\begin{eqnarray}
C_{5L}^{ijkl} &=& (Y_D)_{ij} (V^T P Y_U V)_{kl}, \nonumber \\
C_{5R}^{ijkl} &=& (P^*V^*Y_D)_{ij}(Y_UV)_{kl},
\label{eqn:C5L&C5R}
\end{eqnarray}
where $Y_U$ and $Y_D$ are diagonalized Yukawa coupling matrices 
for $10 \cdot 10 \cdot 5_H$ and 
$10 \cdot \overline{5} \cdot \overline{5}_H$ interactions, respectively. 
The unitary matrix $V$ is the Cabibbo-Kobayashi-Maskawa (CKM) matrix 
at the GUT scale.
The matrix $P$ $=$ ${\rm diag}(P_1,P_2,P_3)$ is a 
diagonal unimodular phase matrix with $|P_i|=1$ and $ {\rm det}P=1$. 
We parametrize $P$ as 
\begin{eqnarray}
P_1/P_3 = e^{i\phi_{13}}, \hspace{3mm}
P_2/P_3 = e^{i\phi_{23}}.
\label{eqn:Phase_Matrix}
\end{eqnarray}
The parameters $\phi_{13}$ and $\phi_{23}$ are relative phases between 
the Yukawa couplings at the colored Higgs interactions, and 
cannot be removed by field redefinitions \cite{Phase_Matrix}. 
The expressions for $C_{5L}$ and $C_{5R}$ in Eq.~(\ref{eqn:C5L&C5R}) 
are written in the flavor basis where the Yukawa coupling matrix for 
the $10 \cdot \overline{5} \cdot \overline{5}_H$ interaction 
is diagonalized at the GUT scale. 
Numerical values of $Y_U$, $Y_D$ and $V$ at the GUT scale are calculated 
from the quark masses and the CKM matrix at the weak scale 
using renormalization group equations (RGEs).

In this model, soft SUSY breaking parameters at the 
Planck scale are described by $m_0$, $M_{gX}$ and $A_X$
which denote universal scalar mass, universal gaugino mass, 
and universal coefficient of the trilinear scalar couplings, respectively. 
Low energy values of the soft breaking parameters are 
determined by solving the one-loop RGEs \cite{Radiative_Breaking}. 
The electroweak symmetry is broken radiatively 
due to the effect of a large Yukawa coupling of the top quark,
and we require that the correct vacuum expectation values of 
the Higgs fields at the weak scale are reproduced. 
Thus we have all the values of the parameters at the weak scale. 
The masses and the mixings are obtained by diagonalizing
the mass matrices numerically. 
We evaluate hadronic matrix elements using the chiral Lagrangian 
method \cite{Chiral_Lagrangian}.
The parameters $\alpha_p$ and $\beta_p$ defined by
$\langle 0| \epsilon^{abc} (d_R^a u_R^b) u_L^c |p 
\rangle$ $=$ $\alpha_p N_L$ and 
$\langle 0| \epsilon^{abc} (d_L^a u_L^b) u_L^c |p 
\rangle$ $=$ $\beta_p N_L$ 
($N_L$ is a wave function of the left-handed proton) 
are evaluated as $0.003 \, {\rm GeV}^3$ $\leq$ $\beta_p$
$\leq$ $0.03 \, {\rm GeV}^3$ and $\alpha_p$ $=$ $-\beta_p$ 
by various methods \cite{beta_p}. 
We use the smallest value 
$\beta_p$ $=$ $-\alpha_p$ $=$ $0.003 \, {\rm GeV}^3$ 
in our analysis to obtain conservative bounds. 
For the details of the methods of our analysis, see Ref.~\cite{GN}. 

%
%%%%%%%%%%%%%%%%%%%%%%%%
%
\section{$RRRR$ contribution to the proton decay}
\begin{figure}[t]
\unitlength 1mm
\epsfxsize=7cm
\hspace*{20mm}
\leavevmode\epsffile{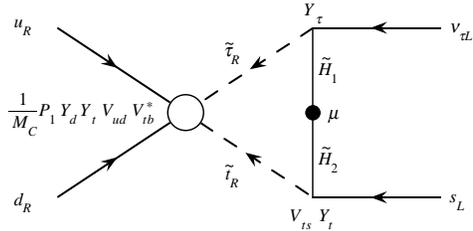}
\caption[figI]{Higgsino dressing diagram which gives a dominant
contribution to the $p\rightarrow K^+ \overline{\nu}_\tau$ mode. 
The circle represents
the $RRRR$ dimension 5 operator. We also have a similar diagram 
for $(u_R s_R)(d_L \nu_{\tau L})$.}
\label{fig:diagram}
\end{figure}

%
%%%%%<<< At M_W >>>%%%%%
%
The dimension 5 operators consist of two fermions and two bosons.
Eliminating the two scalar bosons by gaugino or Higgsino
exchange (dressing), we obtain the four-fermion interactions
which cause the nucleon decay \cite{DRW,NCA}.
In the one-loop calculations of the dressing diagrams,
we include all the dressing diagrams exchanging
the charginos, the neutralinos and the gluino of the
$LLLL$ and $RRRR$ dimension 5 operators. 
%-- dim 6 --
In addition to the contributions from the dimension 5 operators, 
we include the contributions from dimension 6 operators mediated 
by the heavy gauge boson and the colored Higgs boson. 
Though the effects of the dimension 6 operators ($\sim$ $1/M_X^2$)
are negligibly small for $p\rightarrow K^+\overline{\nu}$, 
these could be important for other decay modes such as 
$p\rightarrow \pi^0 e^+$. 
%
%-------------
The major contribution of the $LLLL$ operator comes from
an ordinary diagram with wino dressing. 
The major contribution of the $RRRR$ operator arises from
a Higgsino dressing diagram depicted in Fig.~\ref{fig:diagram}. 
%
%%%+++++
%
%
%%%+++++
%
The circle in this figure represents the complex conjugation of 
$C_{5R}^{ijkl}$ in Eq.~(\ref{eqn:C5L&C5R}) with $i$ $=$ $j$ $=1$ 
and $k$ $=$ $l$ $=3$. This diagram contains the Yukawa couplings
of the top quark and the tau lepton. 
The importance of this diagram has been pointed out in 
Ref.~\cite{RRRR} in the context of a SUSY SO(10) GUT model. 
The contribution of this diagram has been estimated to be negligible 
or simply ignored in previous analyses in the minimal SU(5) SUSY 
GUT \cite{dim5_op,DRW,NCA,MATS+HMY,HMTY,GNA}. 
In particular, the authors of Ref.~\cite{NCA} calculated the diagram in 
Fig.~\ref{fig:diagram}. However the effect was estimated to be small, 
because the authors assumed a relatively light top quark ($m_t$ $\sim$
50 $\gev$). 
We use an experimental value of the top quark mass,
and show that this diagram indeed gives a significant contribution 
in the case of the minimal SU(5) SUSY GUT model also.

Before we present the results of our numerical calculations, 
we give a rough estimation for the decay amplitudes 
for a qualitative understanding of the results. 
In the actual calculations, however, we make full numerical analyses 
including contributions from all the dressing diagrams as well as 
those from dimension 6 operators.
We also take account of 
various effects such as mixings between the SUSY particles.

Aside from the soft breaking parameter dependence arising from the loop
calculations, relative magnitudes 
between various contributions can be roughly 
understood by the form of the dimension 5 operator 
in Eq.~(\ref{eqn:C5L&C5R}). 
Counting the CKM suppression factors and the Yukawa coupling factors, 
it is easily shown that 
the $RRRR$ contribution to
the four-fermion operators $(u_R d_R)(s_L \nu_{\tau L})$
and $(u_R s_R)(d_L \nu_{\tau L})$ is dominated by a single 
Higgsino dressing diagram
exchanging $ \tilde{t}_R$ (the right-handed scalar top quark) 
and $ \tilde{\tau}_R $ (the right-handed scalar tau lepton). 
For $K^+\overline{\nu}_\mu$ and $K^+\overline{\nu}_e$,
the $RRRR$ contribution is negligible, 
since it is impossible to get a large Yukawa coupling
of the third generation without small CKM suppression
factors in this case. 
The main $LLLL$ contributions to $(u_L d_L)(s_L \nu_{i L})$
and $(u_L s_L)(d_L \nu_{i L})$ consist of two 
classes of wino dressing diagrams; they are
$ \tilde{c}_L$ exchange diagrams and $ \tilde{t}_L$ exchange
diagrams \cite{NCA}. 
Neglecting other subleading effects, 
we can write the amplitudes
(the coefficients 
of the four-fermion operators) for 
$p\rightarrow K^+ \overline{\nu}_i$ as, 
\begin{eqnarray}
{\rm Amp.}(p\rightarrow K^+ \overline{\nu}_e) & \sim & 
[ P_2 A_e(\tilde{c}_L) +
  P_3 A_e(\tilde{t}_L) ]_{LLLL}, \nonumber \\
{\rm Amp.}(p\rightarrow K^+ \overline{\nu}_\mu) & \sim & 
[ P_2 A_\mu(\tilde{c}_L) +
  P_3 A_\mu(\tilde{t}_L) ]_{LLLL}, \nonumber \\
{\rm Amp.}(p\rightarrow K^+ \overline{\nu}_\tau) & \sim & 
[ P_2 A_\tau(\tilde{c}_L) +
  P_3 A_\tau(\tilde{t}_L) ]_{LLLL}
+ [ P_1 A_\tau(\tilde{t}_R) ]_{RRRR}, 
\label{eqn:rough_estimation}
\end{eqnarray}
where the subscript $LLLL$ ($RRRR$) represents the contribution
from the $LLLL$ ($RRRR$) operator. 
The $LLLL$ contributions for $A_\tau$ 
can be written in a rough approximation as 
$A_\tau(\tilde{c}_L)$ $\sim$ $g^2 Y_c Y_b 
V_{ub}^* V_{cd}V_{cs}m_{\tilde{W}}$
$\!\! /(M_C m_{\tilde{f}}^2)$ and 
$A_\tau(\tilde{t}_L)$ $\sim$ $g^2 Y_t Y_b
V_{ub}^* V_{td}V_{ts}m_{\tilde{W}}$ 
$\!\! /(M_C m_{\tilde{f}}^2)$,
where $g$ is the weak SU(2) gauge coupling, and 
$m_{\tilde{W}}$ is a mass of the wino $\tilde{W}$. 
A typical mass scale of the squarks and the sleptons 
is denoted by $m_{\tilde{f}}$. 
For $A_\mu$ and $A_e$, we just replace $Y_b V_{ub}^*$ in the
expressions for $A_\tau$ by $Y_s V_{us}^*$ and $Y_d V_{ud}^*$,
respectively. 
The $RRRR$ contribution is also evaluated as 
$A_\tau(\tilde{t}_R)$ $\sim$
$Y_d Y_t^2 Y_\tau V_{tb}^* V_{ud}V_{ts}\mu$
$\!\! /(M_C m_{\tilde{f}}^2)$,
where $\mu$ is the supersymmetric Higgsino mass. 
The magnitude of $\mu$ is determined from the 
radiative electroweak symmetry breaking condition, 
and satisfies $|\mu|$ $>$ $|m_{\tilde{W}}|$ 
in the present scenario. 
Relative magnitudes between these contributions are 
evaluated as follows. 
The magnitude of the $\tilde{c}_L$ contribution 
is comparable with that of the $\tilde{t}_L$ contribution 
for each generation mode: 
$|A_i(\tilde{c}_L)|$ $\sim$ $|A_i(\tilde{t}_L)|$. 
Therefore, cancellations between 
the $LLLL$ contributions 
$P_2 A_i(\tilde{c}_L)$ and $P_3 A_i(\tilde{t}_L)$ 
can occur simultaneously for three modes 
$p\rightarrow K^+\overline{\nu}_i$ ($i$ $=$ $e$, $\mu$ and $\tau$) 
by adjusting the relative phase $\phi_{23}$ 
between $P_2$ and $P_3$ \cite{NCA}. 
The magnitudes of the $LLLL$ contributions satisfy
$|P_2 A_\mu(\tilde{c}_L) + P_3 A_\mu(\tilde{t}_L)|$ 
$>$ $|P_2 A_\tau(\tilde{c}_L) + P_3 A_\tau(\tilde{t}_L)|$
$>$ $|P_2 A_e(\tilde{c}_L) + P_3 A_e(\tilde{t}_L)|$ 
independent of $\phi_{23}$. 
On the other hand, 
the magnitude of $A_\tau(\tilde{t}_R)$ is larger than 
those of $A_i(\tilde{c}_L)$ and $A_i(\tilde{t}_L)$, 
and the phase dependence of $P_1 A_\tau(\tilde{t}_R)$ is 
different from those of 
$P_2 A_i(\tilde{c}_L)$ and $P_3 A_i(\tilde{t}_L)$. 
Note that $A_i(\tilde{c}_L)$ and $A_i(\tilde{t}_L)$
are proportional to $\sim$ $1/(\sin \beta \cos \beta)$ $=$ 
$\tan \beta +  1/\tan \beta$, 
while $A_\tau(\tilde{t}_R)$ 
is proportional to $\sim$ $(\tan \beta +  1/\tan \beta)^2$,
where $\tan \beta$ is the ratio of the vacuum expectation values 
of the two Higgs fields.
Hence the $RRRR$ contribution is more enhanced than the $LLLL$ 
contributions for a large $\tan \beta$ \cite{RRRR}.

%
%%%%%%%%%%%%%%%%%%%%%%%%%%%%%%%%%%%%%%%%%%%%%%%%%%%%%%%%%%%%%%%%%%%%%%%%%%
%
\section{Numerical results}
Now we present the results of our numerical calculations \cite{GN}. 
For the CKM matrix we fix the parameters as 
$V_{us}=0.2196$, $V_{cb}=0.0395$, $|V_{ub}/V_{cb}|=0.08$
and $\delta_{13}=90^\circ$ in the whole analysis,
where $\delta_{13}$ is a complex phase in the CKM matrix 
in the convention of Ref. \cite{standard-parametrization}. 
The top quark mass is taken to be 175 $\gev$ \cite{m_top}. 
The colored Higgs mass $M_C$ and the heavy gauge boson mass
$M_V$ are assumed as $M_C$ $=$ $M_V$ $=$ $2 \times 10^{16} \gev$. 
We require 
constraint on $ b \rightarrow s \gamma $ branching ratio 
from CLEO \cite{CLEO} 
and bounds on SUSY particle masses obtained from 
direct searches at LEP \cite{Neutralino_Bound}, 
LEP II \cite{LEP2} and Tevatron \cite{CDF+D0}.

\begin{figure}[t]
\hspace*{10mm}
\unitlength 1mm
\epsfxsize=8.5cm
\leavevmode\epsffile{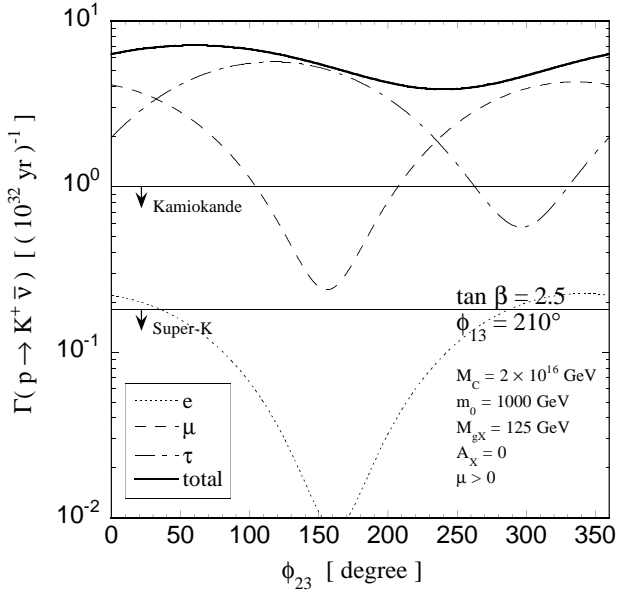}
\caption[figII]{Decay rates
$\Gamma(p\rightarrow K^+ \overline{\nu}_i)$ ($i$ $=$ $e$, $\mu$ and $\tau$) 
as functions of the phase $\phi_{23}$ for $\tan \beta$ $=2.5$ \cite{GN}. 
The other phase $\phi_{13}$ 
is fixed at $210^\circ$. The CKM phase is taken as $\delta_{13}=90^\circ$. 
We fix the soft SUSY breaking parameters
as $m_0$ $=$ $1 \tev$, $M_{gX}$ $=$ $125 \gev$ and $A_X$ $=0$.
The sign of the supersymmetric Higgsino mass $\mu$ 
is taken to be positive. 
The colored Higgs mass $M_C$ and the heavy gauge boson mass
$M_V$ are assumed as $M_C$ $=$ $M_V$ $=$ $2 \times 10^{16} \gev$.
The horizontal lower line corresponds to 
the Super-Kamiokande limit  
$\tau(p\rightarrow K^+ \overline{\nu})$
$>$ $5.5 \times 10^{32}$ years, 
and the horizontal upper line corresponds to the Kamiokande limit 
$\tau(p\rightarrow K^+ \overline{\nu})$ 
$>$ $1.0 \times 10^{32}$ years.}
\label{fig:phi23}
\end{figure}

Let us focus on the main decay mode 
$p\rightarrow K^+\overline{\nu}$. 
We first discuss the effects of the phases $\phi_{13}$ and $\phi_{23}$
parametrizing the matrix $P$ in Eq.~(\ref{eqn:Phase_Matrix}).
In Fig.~\ref{fig:phi23} we present the dependence of 
the decay rates $\Gamma(p\rightarrow K^+ \overline{\nu}_i)$ 
on the phase $\phi_{23}$.  
%
%%%+++++
%
%
%%%+++++
%
As an illustration we fix the other phase $\phi_{13}$ at $210^\circ$, 
and later we consider 
the whole parameter space of $\phi_{13}$ and $\phi_{23}$. 
The soft SUSY breaking parameters are also fixed 
as $m_0$ $=$ $1 \tev$, $M_{gX}$ $=$ $125 \gev$ and $A_X$ $=0$ here.
The sign of the Higgsino mass $\mu$ 
is taken to be positive. 
With these parameters, 
all the masses of the scalar fermions  
other than the lighter $\tilde{t}$ are around 1 $\tev$, 
and the mass of the lighter $\tilde{t}$ is about 400 $\gev$. 
The lighter chargino is wino-like with a mass about 100 $\gev$. 
This figure shows that there is no region for 
the total decay rate $\Gamma(p\rightarrow K^+ \overline{\nu})$
to be strongly suppressed, thus the whole region of $\phi_{23}$ 
in Fig.~\ref{fig:phi23} is excluded by the Super-Kamiokande limit.
The phase dependence of 
$\Gamma(p\rightarrow K^+ \overline{\nu}_\tau)$ is 
quite different from those of 
$\Gamma(p\rightarrow K^+ \overline{\nu}_\mu)$ 
and $\Gamma(p\rightarrow K^+ \overline{\nu}_e)$. 
Though $\Gamma(p\rightarrow K^+ \overline{\nu}_\mu)$ 
and $\Gamma(p\rightarrow K^+ \overline{\nu}_e)$
are highly suppressed around $\phi_{23}$ $\sim$ $160^\circ$, 
$\Gamma(p\rightarrow K^+ \overline{\nu}_\tau)$ is not so 
in this region. 
There exists also the region $\phi_{23}$ $\sim$ $300^\circ$ where 
$\Gamma(p\rightarrow K^+ \overline{\nu}_\tau)$ is reduced. 
In this region, however, $\Gamma(p\rightarrow K^+ \overline{\nu}_\mu)$ 
and $\Gamma(p\rightarrow K^+ \overline{\nu}_e)$ 
are not suppressed in turn. 
Note also that the $K^+\overline{\nu}_\tau$ mode
can give the largest contribution.

This behavior can be understood as follows. 
For $\overline{\nu}_\mu$ and $\overline{\nu}_e$, the effect of
the $RRRR$ operator is negligible, and 
the cancellation between the $LLLL$ contributions 
directly leads to the suppression of the decay rates. 
This cancellation indeed occurs around 
$\phi_{23}$ $\sim$ $160^\circ$ for 
both $\overline{\nu}_\mu$ 
and $\overline{\nu}_e$ simultaneously in Fig.~\ref{fig:phi23}. 
For $\overline{\nu}_\tau$, the situation is quite different.
The similar cancellation between 
$P_2 A_\tau(\tilde{c}_L)$ and $P_3 A_\tau(\tilde{t}_L)$ 
takes place around $\phi_{23}$ $\sim$ $160^\circ$ for 
$\overline{\nu}_\tau$ also. 
However, the $RRRR$ operator gives 
a significant contribution for $\overline{\nu}_\tau$. 
Therefore, $\Gamma(p\rightarrow K^+ \overline{\nu}_\tau)$
is not suppressed by the cancellation between the $LLLL$ contributions 
in the presence of the large $RRRR$ operator effect.
Notice that it is possible 
to reduce $\Gamma(p\rightarrow K^+\overline{\nu}_\tau)$ 
by another cancellation between the $LLLL$ contributions 
and the $RRRR$ contribution. 
This reduction of 
$\Gamma(p\rightarrow K^+\overline{\nu}_\tau)$ 
indeed appears around 
$\phi_{23}$ $\sim$ $300^\circ$ in Fig.~\ref{fig:phi23}. 
The decay rate $\Gamma(p\rightarrow K^+\overline{\nu}_\mu)$ is 
rather large in this region. 
The reason is that 
$P_2 A_\tau(\tilde{c}_L)$ and $P_3 A_\tau(\tilde{t}_L)$ 
are constructive in this 
region in order to cooperate with each other 
to cancel the large $RRRR$ contribution 
$P_1 A_\tau(\tilde{t}_R)$, hence 
$P_2 A_\mu(\tilde{c}_L)$ and $P_3 A_\mu(\tilde{t}_L)$
are also constructive in this region. 
Thus we cannot reduce both $\Gamma(p\rightarrow K^+\overline{\nu}_\tau)$
and $\Gamma(p\rightarrow K^+\overline{\nu}_\mu)$ simultaneously. 
Consequently, there is no region for 
the total decay rate $\Gamma(p\rightarrow K^+ \overline{\nu})$
to be strongly suppressed. 
%
%-------------
In the previous analysis \cite{HMTY} the region 
$\phi_{23}$ $\sim$ $160^\circ$ has been considered to be allowed 
by the Kamiokande limit $\tau(p\rightarrow K^+ \overline{\nu})$ 
$>$ $1.0 \times 10^{32}$ years (90\,\% C.L.) \cite{Kam}. 
However the inclusion of the Higgsino dressing of the 
$RRRR$ operator excludes this region. 
%
%-------------
%
We also examined the whole region of $\phi_{13}$ and $\phi_{23}$
with the same values for the other parameters as in Fig.~\ref{fig:phi23}. 
We found that we cannot reduce 
both $\Gamma(p\rightarrow K^+\overline{\nu}_\tau)$
and $\Gamma(p\rightarrow K^+\overline{\nu}_\mu)$ simultaneously, 
even if we adjust the two phases $\phi_{13}$ and $\phi_{23}$ 
anywhere.

\begin{figure}[t]
\hspace*{10mm}
\unitlength 1mm
\epsfxsize=8.5cm
\leavevmode\epsffile{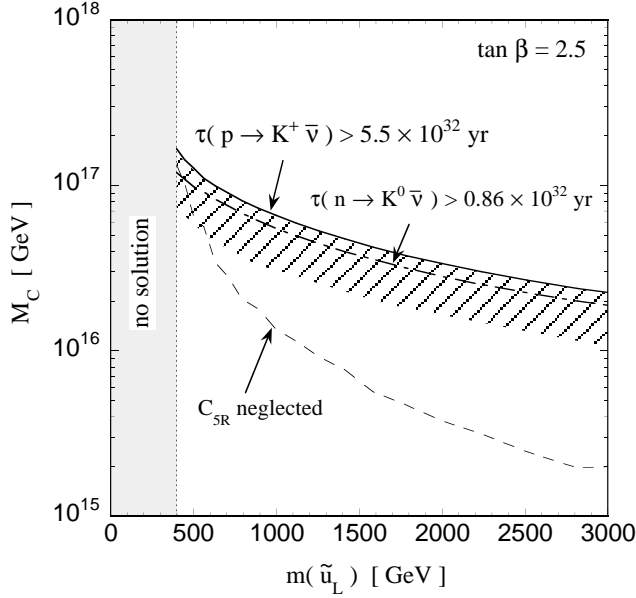}
\caption[figIV]{Lower bound on the colored Higgs mass $M_C$
as a function of the 
left-handed scalar up-quark mass $m_{\tilde{u}_L}$ \cite{GN}. 
The soft breaking parameters $m_0$, $M_{gX}$ and $A_X$ are scanned 
within the range of $0<m_0<3 \tev$, $0<M_{gX}<1 \tev$ and $ -5<A_X<5 $, 
and $\tan \beta$ is fixed at 2.5.
Both signs of $\mu$ are considered. 
The whole parameter region of the two phases $\phi_{13}$ and $\phi_{23}$ 
is examined. 
The solid curve represents the bound derived from the 
Super-Kamiokande limit $\tau(p\rightarrow K^+ \overline{\nu})$
$>$ $5.5 \times 10^{32}$ years, and 
the dashed curve represents the corresponding result 
without the $RRRR$ effect. 
The left-hand side of the vertical dotted line is excluded by 
other experimental constraints.
The dash-dotted curve represents the bound 
derived from the Kamiokande limit on the neutron partial lifetime 
$\tau (n\rightarrow K^0 \overline{\nu})$ 
$>$ $0.86 \times 10^{32}$ years.}
\label{fig:m_sf}
\end{figure}

Next we consider the case where 
we vary the parameters we have fixed so far. 
The relevant parameters are 
the colored Higgs mass $M_C$, 
the soft SUSY breaking parameters and $\tan \beta$.  
The partial lifetime $\tau(p\rightarrow K^+ \overline{\nu})$ 
is proportional to $M_C^2$ in a very good approximation,
since this mode is dominated by the dimension 5 operators. 
Using this fact and the calculated value of 
$\tau(p\rightarrow K^+ \overline{\nu})$ for the fixed 
$M_C$ $=$ $2 \times 10^{16} \gev$, we can obtain the lower bound
on $M_C$ from the experimental lower limit 
on $\tau(p\rightarrow K^+ \overline{\nu})$. 
In Fig.~\ref{fig:m_sf}, we present 
the lower bound on $M_C$ obtained from 
the Super-Kamiokande limit as a function of the 
left-handed scalar up-quark mass $m_{\tilde{u}_L}$. 
%
%%%+++++
%
%
%%%+++++
%
Masses of the squarks other than the lighter $\tilde{t}$ 
are almost degenerate with $m_{\tilde{u}_L}$. 
The soft breaking parameters $m_0$, $M_{gX}$ and $A_X$ are scanned 
within the range of $0<m_0<3 \tev$, $0<M_{gX}<1 \tev$ and $ -5<A_X<5 $, 
and $\tan \beta$ is fixed at 2.5.
Both signs of $\mu$ are considered. 
The whole parameter region of the two phases $\phi_{13}$ and $\phi_{23}$ 
is examined. 
The solid curve in this figure 
represents the result with all the $LLLL$ and $RRRR$ contributions. 
It is shown that the lower bound on $M_C$ decreases 
like $\sim$ $1/m_{\tilde{u}_L}$ as $m_{\tilde{u}_L}$ increases. 
This indicates that the $RRRR$ effect is indeed relevant,
since the decay amplitude from the $RRRR$ operator is roughly proportional
to $\mu/(M_C m_{\tilde{f}}^2)$ $\sim$ $1/(M_C m_{\tilde{f}})$, 
where we use the fact that the magnitude of $\mu$ is determined from the
radiative electroweak symmetry breaking condition and
scales as $\mu$ $\sim$ $m_{\tilde{f}}$. 
The dashed curve in Fig.~\ref{fig:m_sf} 
represents the result in the case 
where we ignore the $RRRR$ effect. 
In this case 
the lower bound on $M_C$ behaves as 
$\sim$ $1/m_{\tilde{u}_L}^2$, since the $LLLL$ contribution 
is proportional to $m_{\tilde{W}}/(M_C m_{\tilde{f}}^2)$.

The solid curve in Fig.~\ref{fig:m_sf} indicates 
that the colored Higgs mass $M_C$ must be larger 
than $6.5 \times 10^{16} \gev$ for $\tan \beta$ $=$ 2.5 when 
the typical sfermion mass is less than $1 \tev$. 
%
%**********
On the other hand, there is a theoretical upper bound of the 
colored Higgs mass $M_C$ $\lsim$ $4 \times 10^{16} \gev$ from 
an analysis of RGEs 
when we require the validity of the minimal SU(5) SUSY GUT model 
up to the Planck scale \footnote{
Also it has been pointed out that there exists an upper bound 
on $M_C$ given by $M_C$ $\leq$ $2.5 \times 10^{16} \gev$ (90\,\% C.L.) if 
we require the gauge coupling unification in the minimal contents of GUT 
superfields \cite{HMTY}. }. 
Then it follows from Fig.~\ref{fig:m_sf} that the minimal SU(5) SUSY GUT 
model with the sfermion masses less than $2 \tev$ is excluded for 
$\tan \beta$ $=2.5$. 
The $RRRR$ effect plays an essential role here, 
since the lower bound on $m_{\tilde{f}}$ would be $600 \gev$ 
if the $RRRR$ effect were ignored. 
%**********
%
We also find that 
the Kamiokande limit on the neutron partial lifetime 
$\tau (n\rightarrow K^0 \overline{\nu})$ 
$>$ $0.86 \times 10^{32}$ years (90\,\% C.L.) \cite{Kam} 
already gives a comparable bound with that derived here from 
the Super-Kamiokande limit on 
$\tau (p\rightarrow K^+ \overline{\nu})$, 
as shown by the dash-dotted curve in Fig.~\ref{fig:m_sf}.

\begin{figure}[t]
\hspace*{10mm}
\unitlength 1mm
\epsfxsize=8.5cm
\leavevmode\epsffile{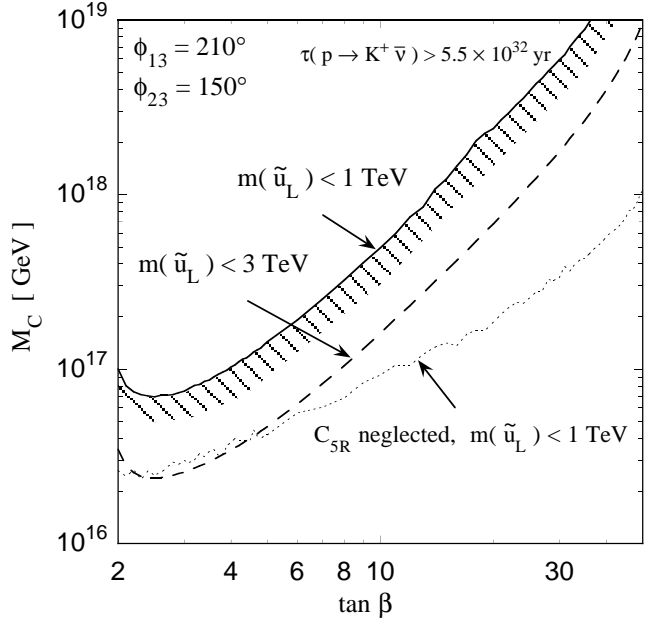}
\caption[figV]{The lower bound 
on the colored Higgs mass $M_C$ obtained from the Super-Kamiokande limit 
as a function of $\tan \beta$ \cite{GN}. 
The phase matrix $P$ is fixed by $\phi_{13}$ $=210^\circ$ 
and $\phi_{23}$ $=150^\circ$.
The region below the solid curve is excluded if the 
left-handed scalar up-quark mass $m_{\tilde{u}_L}$ is less than $1 \tev$. 
The lower bound reduces to the dashed curve if we 
allow $m_{\tilde{u}_L}$ up to $3 \tev$.
The result in the case where we ignore the $RRRR$ effect 
is shown by the dotted curve for $m_{\tilde{u}_L}$ $<$ $1 \tev$.}
\label{fig:tanB}
\end{figure}

Fig.~\ref{fig:tanB} shows the $\tan \beta$ dependence of the lower bound 
on the colored Higgs mass $M_C$ obtained from the Super-Kamiokande limit.
%
%%%+++++
%
%
%%%+++++
%
Here we vary $m_0$, $M_{gX}$ and $A_X$ as in Fig.~\ref{fig:m_sf}.
The phases $\phi_{13}$ and $\phi_{23}$ are fixed as 
$\phi_{13}$ $=210^\circ$ and $\phi_{23}$ $=150^\circ$. 
The result does not change much even if we take other values of 
$\phi_{13}$ and $\phi_{23}$. 
The region below the solid curve is excluded if 
$m_{\tilde{u}_L}$ is less than $1 \tev$. 
The lower bound reduces to the dashed curve if we 
allow $m_{\tilde{u}_L}$ up to $3 \tev$. 
It is shown that the lower bound on $M_C$ 
behaves as $\sim$ $\tan^2 \beta$ in 
a large $\tan \beta$ region, as expected from the fact that 
the amplitude of $p\rightarrow K^+ \overline{\nu}_\tau$ 
from the $RRRR$ operator is roughly 
proportional to $\tan^2 \beta /M_C$. 
On the other hand the $LLLL$ contribution is 
proportional to $\sim$ $\tan \beta /M_C$, 
as shown by the dotted curve in Fig.~\ref{fig:tanB}. 
Thus the $RRRR$ operator is dominant for large $\tan \beta$ \cite{RRRR}. 
Note that the lower bound on $M_C$ has the minimum at 
$\tan \beta$ $\approx$ 2.5. 
Thus we can conclude that for other value of $\tan \beta$ 
the constraints on $M_C$ and $m_{\tilde{f}}$ become severer 
than those shown in Fig.~\ref{fig:m_sf}. 
In particular the lower bound on $m_{\tilde{f}}$ becomes larger 
than $\sim$ $2 \tev$ for $\tan \beta$ $\neq$ 2.5. 

The constraints obtained from the figures can be expressed as follows: 
\begin{eqnarray}
\left( \frac{M_C}{6.5 \times 10^{16} \gev}\right) & \gsim & 
\left( \frac{\tau^{\rm exp}(p \rightarrow K^{+} \overline{\nu})}
     {5.5 \times 10^{32} {\rm years}} \right)^{\frac{1}{2}}
\left( \frac{\beta_p}{0.003 \gev^3} \right)
\left( \frac{1 \tev}{m_{\tilde{f}}} \right)
 \nonumber \\
 & & \hspace{3cm} {\rm for} \ \tan \beta \approx 2.5, \nonumber \\
\left( \frac{M_C}{5.0 \times 10^{17} \gev}\right) & \gsim & 
\left( \frac{\tau^{\rm exp}(p \rightarrow K^{+} \overline{\nu})}
     {5.5 \times 10^{32} {\rm years}} \right)^{\frac{1}{2}}
\left( \frac{\beta_p}{0.003 \gev^3} \right)
\left( \frac{1 \tev}{m_{\tilde{f}}} \right)
\left( \frac{\tan \beta}{10} \right)^2
\nonumber \\
 & & \hspace{3cm} {\rm for} \ \tan \beta \ \gsim \ 5, 
\label{eqn:constraint}
\end{eqnarray}
where $\tau^{\rm exp}(p \rightarrow K^{+} \overline{\nu})$ is an experimental
lower limit for the partial lifetime of the decay mode 
$p \rightarrow K^{+} \overline{\nu}$. 

%
%%%%%%%%%%%%%%%%%%%%%%%%%%%%%%%%%%%%%%%%%%%%%%%%%%%%%%%%%%%%%%%%%%%%%%%%%%
%
\section{Conclusions}
We have reanalyzed the proton decay 
including the $RRRR$ dimension 5 operator 
in the minimal SU(5) SUSY GUT model.
We have shown that the Higgsino dressing diagram of 
the $RRRR$ operator gives a dominant
contribution for 
$p\rightarrow K^+\overline{\nu}_\tau$,
and the decay rate of this mode 
can be comparable with that of 
$p\rightarrow K^+\overline{\nu}_\mu$. 
We have found that 
we cannot reduce both the decay rate of 
$p\rightarrow K^+\overline{\nu}_\tau$ and 
that of $p\rightarrow K^+\overline{\nu}_\mu$ simultaneously
by adjusting the relative phases $\phi_{13}$ and $\phi_{23}$ 
between the Yukawa couplings
at the colored Higgs interactions. 
We have obtained the bounds on the colored Higgs mass $M_C$ 
and the typical sfermion mass $m_{\tilde{f}}$ from 
the new limit on $\tau(p\rightarrow K^+ \overline{\nu})$ given by 
the Super-Kamiokande: 
$M_C$ $>$ $6.5 \times 10^{16} \gev$ for $m_{\tilde{f}}$ $<$ $1 \tev$.
The minimal SU(5) SUSY GUT model with $m_{\tilde{f}}$ $\lsim$ $2 \tev$ is 
excluded if we require the validity of this model up to the Planck scale. 
%

%
%%%%%%%%%%%%%%%%%%%%%%%%%%%%%%%%%%%%%%%%%%%%%%%%%%%%%%%%%%%%%%%%%%%%%%%%%%
%
%\section*{Acknowledgments}
%This is where one places acknowledgments for funding
%bodies etc.  Note that there are no section numbers for
%the Acknowledgments, Appendix or References.
%
%
%%%%%%%%%%%%%%%%%%%%%%%%%%%%%%%%%%%%%%%%%%%%%%%%%%%%%%%%%%%%%%%%%%%%%%%%%%
%


\section*{References}
\begin{thebibliography}{99}
%
%----- GN -----
%
\bibitem{GN}
T. Goto and T. Nihei, \Journal{\PRD}{59}{115009}{1999}.
%
%
%----- SUSY GUT -----
%
\bibitem{SUSY_GUT}
E. Witten, \Journal{\NPB}{188}{513}{1981}; 
S. Dimopoulos, S. Raby and F. Wilczek, \Journal{\PRD}{24}{1681}{1981}; 
S. Dimopoulos and H. Georgi, \Journal{\NPB}{193}{150}{1981}; 
N. Sakai, \Journal{\ZPC}{11}{153}{1981}.
%
%----- gauge coupling unification -----
%
\bibitem{Gauge_Coupling_Unification}
C. Giunti, C.W. Kim and U.W. Lee, \Journal{\MPL}{6}{1745}{1991}; 
J. Ellis, S. Kelley and D.V. Nanopoulos, \Journal{\PLB}{260}{131}{1991};
U. Amaldi, W. de Boer and H. F\"{u}rstenau, \Journal{\PLB}{260}{447}{1991};
P. Langacker and M.-X. Luo, \Journal{\PRD}{44}{817}{1991}; 
W.J. Marciano, {\em Ann. Rev. Nucl. Part.} {\bf 41}, 469 (1991). 
%
%------
\bibitem{DRW}
S. Dimopoulos, S. Raby and F. Wilczek, \Journal{\PLB}{112}{133}{1982}; 
J. Ellis, D.V. Nanopoulos and S. Rudaz, \Journal{\NPB}{202}{43}{1982}.
%
%----- SUGRA -----
%
\bibitem{SUGRA}
For reviews on the minimal SU(5) SUGRA GUT model, see for instance, 
H.P. Nilles, \Journal{\PR}{110}{1}{1984};
P. Nath, R. Arnowitt and A.H. Chamseddine, 
Applied $N=1$ Supergravity (World Scientific, Singapore, 1984). 
%
%------- EXPERIMENTS -------
\bibitem{Kam}
Kamiokande Collaboration, K.S. Hirata {\it et al.}, 
\Journal{\PLB}{220}{308}{1989}.
%
%-----
\bibitem{IMB}
IMB Collaboration, R. Becker-Szendy {\it et al.}, 
Proceedings of 23rd International 
Cosmic Ray Conference, Calgary 1993 {\bf Vol.4} 589.
%
%-----
\bibitem{superK}
M. Takita (Super-Kamiokande Collaboration),
Talk presented in 29th International Conference on High Energy 
Physics, Vancouver, July 1998.
%
%-----
\bibitem{superK-new}
Super-Kamiokande Collaboration, \Journal{\PRL}{83}{1529}{1999}.
%
%
%----- pdecay -----
%
\bibitem{dim5_op}
N. Sakai and T. Yanagida, \Journal{\NPB}{197}{533}{1982};
S. Weinberg, \Journal{\PRD}{26}{287}{1982}.
%---
\bibitem{NCA}
P. Nath, A.H. Chamseddine and R. Arnowitt, \Journal{\PRD}{32}{2348}{1985}.
%
%---
\bibitem{MATS+HMY}
M. Matsumoto, J. Arafune, H. Tanaka and K. Shiraishi, 
\Journal{\PRD}{46}{3966}{1992};
J. Hisano, H. Murayama and T. Yanagida, \Journal{\NPB}{402}{46}{1993}.
%
\bibitem{HMTY}
J. Hisano, T. Moroi, K. Tobe and T. Yanagida, \Journal{\MPL}{10}{2267}{1995}.
%
\bibitem{GNA}
T. Goto, T. Nihei and J. Arafune, \Journal{\PRD}{52}{505}{1995}.
%
%
%----- RRRR -----
\bibitem{RRRR}
V. Lucas and S. Raby, \Journal{\PRD}{55}{6986}{1997}.
%
%----- phase matrix -----
\bibitem{Phase_Matrix}
J. Ellis, M.K. Gaillard and D.V. Nanopoulos, \Journal{\PLB}{88}{320}{1979}.
%
%
%----- radiative breaking -----
%
\bibitem{Radiative_Breaking}
K. Inoue, A. Kakuto, H. Komatsu and S. Takeshita, 
\Journal{\PTP}{68}{927}{1982}; {\it ibid.}\ {\bf 71}, 413 (1984); 
L. Ib\'{a}\~{n}ez and G.G. Ross, \Journal{\PLB}{110}{215}{1982};
L. Alvarez-Gaum\'{e}, J. Polchinski and M.B. Wise,
\Journal{\NPB}{221}{495}{1983}; 
J. Ellis, J.S. Hagelin, D.V. Nanopoulos and K. Tamvakis,
\Journal{\PLB}{125}{275}{1983}.
A. Bouquet, J. Kaplan, and C.A. Savoy, 
\Journal{\PLB}{148}{69}{1984}; \Journal{\NPB}{262}{299}{1985}.
%
%--- chiral Lagrangian ---
\bibitem{Chiral_Lagrangian}
M. Claudson, M.B. Wise and L.J. Hall, \Journal{\NPB}{195}{297}{1982}; 
S. Chadha and M. Daniel, \Journal{\NPB}{229}{105}{1983}.
%
%
\bibitem{beta_p}
S.J. Brodsky, J. Ellis, J.S. Hagelin and C.T. Sachrajda,
\Journal{\NPB}{238}{561}{1984}; 
M.B. Gavela, S.F. King, C.T. Sachrajda, G. Martinelli,
M.L. Paciello and B. Taglienti, \Journal{\NPB}{312}{269}{1989}.
%
%
%----- PDG -----
%
\bibitem{standard-parametrization}
L.-L. Chau and W.-Y. Keung, \Journal{\PRL}{53}{1802}{1984}; 
H. Harari and M. Leurer, \Journal{\PLB}{181}{123}{1986}; 
H. Fritzsch and J. Plankl, \Journal{\PRD}{35}{1732}{1987}; 
F.J. Botella and L.-L. Chao, \Journal{\PLB}{168}{97}{1986}.
%
%
\bibitem{m_top}
CDF Collaboration, F. Abe {\it et al}., 
\Journal{\PRD}{50}{2966}{1994}; \Journal{\PRL}{74}{2626}{1995}; 
D0 Collaboration, S. Abachi {\it et al}., \Journal{\PRL}{74}{2632}{1995}.
%
%
%----- b->s+gamma
\bibitem{CLEO}
CLEO Collaboration, M.S. Alam, {\it et al}., \Journal{\PRL}{74}{618}{1995}.
%
%-----
\bibitem{Neutralino_Bound}
L3 Collaboration, M. Acciarri {\it et al}., \Journal{\PLB}{350}{109}{1995}.
%
%
\bibitem{LEP2}
D. Treille, 
Talk presented in 29th International Conference on High Energy 
Physics, Vancouver, July 1998.
%
%
\bibitem{CDF+D0}
CDF Collaboration, F. Abe {\it et al}.,
\Journal{\PRL}{75}{613}{1995}; {\it ibid.}\ {\bf 69} (1992) 3439; 
D0 Collaboration,  S. Abachi {\it et al}., \Journal{\PRL}{75}{618}{1995}.
%
%
\bibitem{Babu-Strassler}
K.S. Babu and M.J. Strassler, hep-ph/9808447.
%
%
\end{thebibliography}
\end{document}